\documentclass[11pt]{article}
\usepackage{amsmath,epsf,amssymb,latexsym,enumerate,cite,shadow,array,color}
\usepackage[ps,dvips,matrix,arrow,frame,import,curve,color]{xy}
\DeclareFontFamily{U}{rsf}{}
\DeclareFontShape{U}{rsf}{m}{n}{
  <5> <6> rsfs5 <7> <8> <9> rsfs7 <10-> rsfs10}{}
\DeclareMathAlphabet\Scr{U}{rsf}{m}{n}

\makeatletter
\@addtoreset{equation}{section}
\makeatother

\def\be{\begin{equation}}
\def\ee{\end{equation}}

\def\ba{\begin{array}}
\def\ea{\end{array}}

\newcommand{\bea}{\begin{eqnarray}}
\newcommand{\eea}{\end{eqnarray}}
\def\N{$\cal N$}
\def\E {$E_{(7,7)}$}

\def\O{\Scr{O}}

\def\cV{{\Scr V}}

\begin{document}
\hfill{}

\vskip 1cm

\vspace{24pt}

\begin{center}
{ \LARGE {\bf  On a possibility of a UV finite    \\
\vskip 0.8cm

  \N=8 supergravity  }}

\vspace{1.5cm}

  {\large  {\bf   Renata Kallosh}}

    \vspace{15pt}

 {Department of Physics,
    Stanford University, Stanford, CA 94305}

\vspace{10pt}

\vspace{24pt}

\end{center}

\begin{abstract}

The Lorentz covariant all-loop counterterms built in the 4+32  on shell superspace of \N=8 supergravity imply that the theory is not UV finite. Meanwhile, the relevant counterterms depending on the light-cone superfields in 4+16 superspace  have not been constructed so far.  Our first attempt to construct them suggests  that they may be incompatible with the  covariant ones.  This would lead to a prediction  of all-loop  UV finiteness  of the perturbative  S-matrix. Here we rely on the validity of the equivalence theorem for  the   S-matrix in the light-cone gauge and Lorentz  covariant gauges, which  requires the absence of  BRST anomalies. We discuss the status of \N=8 $SU(8)$ and \E \,  anomalies. It remains an outstanding problem to construct the  light-cone counterterms or  to confirm our current conclusion.

\end{abstract}
\newpage

\section{Introduction}

The possibility of all-loop UV finiteness of   \N=8 supergravity \cite{Cremmer:1979up}  in perturbation theory was proposed recently \cite{Bern:2006kd,Bern:2007hh}.  The actual results of the 3-loop computations  \cite{Bern:2007hh} confirm the prediction and raise the tantalizing question about the situation  with UV  divergences in the higher loops. It also  leaves us with the puzzle as to why exactly the linearized 3-loop counterterm constructed in \cite{Kallosh:1980fi,Howe:1981xy} did not support the corresponding logarithmic divergence. One possible explanation relies on the assumption \cite{Howe:2002ui} that  in \N=8 supergravity one can construct\footnote{It has not been  constructed so far.}  a generalization of the known  harmonic superspace \cite{Galperin:1984av}. In such case the UV divergences in d=4 would occur starting at the 6-loop order \cite{Howe:2002ui} or at the 5-loop order \cite{Howe:2004pn, Stelle} and at the 4-loop order in d=5.

 Another explanation is that the 3-loop counterterm is only known in the linearized form which should not  be trusted, for example,  in view of the unclear status of the non-linear $E_{(7,7)}$ symmetry. However, the same linearized superfield construction describes well  the {\it non-local} UV finite manifestly supersymmetric one-loop result \cite{Kallosh:2007ym}. Linearized superfields implement  \N=8 global supersymmetry  Ward identities presented explicitly in appendix E.3 of \cite{Bern:1998ug}, whereas the fully non-linear superfields respect the local non-linear  \N=8 supersymmetry.

Independently  of the situation with the onset of divergences,  many supergravity experts  believe  that, most likely, starting from some loop order  there may be an infinite number of divergences.    According to \cite{Howe:1980th}, \cite{Kallosh:1980fi},
they may start at the 8-loop level. The first full superspace integral over torsions, a candidate for the 8-loop UV divergence,  is given by the following 4+32  superspace integral:
\be
S^{L=8}  \sim {\kappa^{14} \over \epsilon} \int d^4 x \; d^{32} \theta  \; \rm Ber E \; T_{ijk\, \alpha} (x, \theta)  \overline T^{ijk \,\dot \alpha} (x, \theta)  T_{mnl }{}^{ \alpha} (x, \theta) \overline T^{mnl }{}_{\dot \alpha}(x, \theta)  \ .
\label{S8}\ee
Here $T_{ijk \,\alpha}(x, \theta) $ is the superspace torsion superfield depending on 32 Grassmann variables. Its first component is a  spin 1/2 field, $T_{ijk \, \alpha}(x, 0)=\chi_{ijk \,\alpha}(x)$. This is a {\it constrained superfield} satisfying some constraints which makes the geometric superspace consistent and all component fields satisfying a full non-linear set of field equations:
\bea
D_\alpha^l  T_{ijk \,\beta}(x, \theta)&=& 3 \delta^l_{[i} M_{jk]\, \alpha \beta}(x, \theta)\ , \nonumber\\
D_\alpha^l  M_{ij \,\beta \gamma}(x, \theta)&=& 2 \delta^l_{[i} \Psi_{j] \, \alpha \beta \gamma}(x, \theta)\ ,\nonumber\\
D_\alpha^l  \Psi_{i\, \beta \gamma \delta }(x, \theta)&=&  \delta^l_{i} C_{ \alpha \beta \gamma \delta}(x, \theta)\ ,
\eea
where $M_{jk \, \alpha \beta}(x, \theta)$ is the superspace curvature tensor whose first component is the vector field strength spinor,
$M_{jk \, \alpha \beta}(x, 0)= F_{jk\,  \alpha \beta}(x) $. The gravitino field strength is given by $ \Psi_{j \, \alpha \beta \gamma}(x, 0)$ and the curvature Weyl spinor is the first component of the superfield $C_{ \alpha \beta \gamma \delta}(x, \theta)$.

The scalars of \N=8 supergravity are in the  ${E_{(7,7)}\over SU(8)}$ coset space and the spinor torsion superfield is related to the spinorial derivative of the scalar superfield. The corresponding on shell covariant  4+32 dimensional superspace was constructed in  \cite{Brink:1979nt}. The infinite number of geometric superinvariants like the one in eq. (\ref{S8}) have unbroken hidden  $E_{(7,7)}$  symmetry: the superspace torsion tensor  $T_{ijk\, \alpha} (x, \theta) $ is an  $E_{(7,7)}$ invariant, and an $SU(8)$ tensor.

A different description of the on shell \N=8 superspace in the light-cone gauge is available in \cite{Brink:1982pd}, \cite{Brink:2008qc}. Instead of the superspace  torsion and curvatures satisfying constrains,  in the light cone gauge there is an {\it unconstrained superfield } which is CPT invariant and relates only physical states. In a proper chiral basis it depends on 16 $\theta$ 's and does not depend on 16 $\bar \theta$ 's.
\bea
\phi (x, \theta, \bar \theta) = \phi (y, \theta)\ , \qquad y^+= x^+   ,  \quad \vec x=\vec y \ , \quad y^- = x^- - {i\over \sqrt
2} \theta^m \bar \theta_m \ .
\eea
Using the notation of \cite{Brink:2008qc}, which include
\bea
\theta^{a_1...a_n}\equiv {1\over n!} \theta^{a_1}... \theta^{a_2}, \qquad \tilde \theta_{a_1...a_n}\equiv \epsilon_ {a_1...a_n b_1... b_{n-8}} \theta^{b^1... b^{n-8}},
\eea
the unconstrained superfield   linear in physical fields is
\bea
&& \phi(y, \theta)= {1\over \partial ^{+ 2}} h(y) +i\theta^m  {1\over \partial ^{+ 2}}\bar \psi_m(y) +i \theta^{mn}  {1\over \partial ^{+ }}\bar B_{mn}(y) -\theta^{mnp}    {1\over \partial ^{+ }} \bar \chi_{mnp}(y)\nonumber\\
\nonumber\\
&&- \theta^{mnpq}  \phi_{mnpq} (y) +i \tilde \theta_{mnp}   \chi^{mnp} +i \tilde \theta_{mn} \partial ^{+} B^{mn}(y) + \tilde \theta_m \partial ^{+} \psi^m(y) + 4 \tilde \theta \partial ^{+2} \bar h(y) \ .
\label{Phi}\eea
It may be useful to remind that the analogous light-cone superfield in \N=4 super Yang-Mills theory has been used as one of the  proofs of the UV finiteness of the theory in \cite{Mandelstam:1982cb}.  The action of the \N=4 super Yang-Mills theory is known, and the supergraph Feynman rules have been analyzed with the simple conclusion (with account of subtleties of a definition of the inverse to ${p^+}$) operator:  the theory is UV finite. The  all-loop UV finiteness of \N=4 super Yang-Mills theory using covariant  superfields  was proved in \cite{Howe:1982tm}.

In \N=8 supergravity  we do not know the  full action in terms of the unconstrained superfield, only the first terms have been constructed recently \cite{Brink:2008qc}. The procedure is based on the form of the $E_{(7,7)}$ symmetry with  an  infinite number of terms in the  transformations.  Recently a new compact form of  $E_{(7,7)}$ symmetry was  discovered in \cite{Kallosh:2008ic},   so it is possible that combining these ideas may help to construct the full action in the manageable form. In such case one would be able to study the UV properties of \N=8 in a way analogous to \N=4 YM in  \cite{Mandelstam:1982cb}.

Before this program is realized, we may try some shortcuts based on the comparison of the structure of the  UV counterterms in Lorentz covariant versus light-cone gauges and rely on the equivalence theorem.
The main idea of our paper can be explained as follows. If the S-matrix is computed in   Lorentz covariant gauges, one expects the UV divergences of the matrix elements to be deduced  from the countertems in the constrained geometric Lorentz covariant form, as shown for the example of the  8th loop in eq. (\ref{S8}). On the other hand, if the computation is performed in the light-cone gauge, the relevant counterterms have to be constructed using the superfield (\ref{Phi}).

In accordance with the equivalence theorem,  the S-matrix contribution from the candidate counterterms in these two gauges must coincide. Note,  however,
that  without actually performing the higher loop calculations  one can only determine the structure of the candidate counterterms but not the coefficient in front of these structures. If we find a mismatch between these structures, for example, if  in one of these gauges the corresponding counterterm  contribution is Lorentz covariant and in another gauge it is not Lorentz covariant, this will imply that  the coefficient in front of the counterterms  must  vanish.  This would  imply  that the relevant UV divergences are  absent.

Thus the first part of our program is to find   the  contribution  from  the  Lorentz covariant counterterms  to  the UV infinite part of the S-matrix at various loop orders. The next step is to construct the candidate all-loop counterterms in the light-cone superspace which will give the UV infinite part of the S-matrix. And, finally, we will compare the structure of the UV divergences  obtained by these two methods  for each loop order.
As we will see,  assuming that we did not miss any candidate counterterms in the light cone gauge,  there is indeed a mismatch between the contributions of the candidate counterterms to the S-matrix. To resolve the discrepancy,  the candidate UV infinities  must vanish. Of course, this conclusion is  based on the equivalence theorem, which relies on the absence of anomalies, so this issue will still require careful investigation.

\section{The Lorentz covariant counterterms and S-matrix}

We start with the 3-loop type \N=8 superinvariant with the gravitational  part which is a square of the Bel-Robinson tensor \cite{Kallosh:1980fi,Howe:1981xy}.
For the 4-particle case we are interested in the linearized part of the following expression\footnote{ This expression as a  candidate for the 3-loop divergence in \N=1 supergravity  was proposed in \cite{Deser:1977nt}  and generalized  to \N=8 in \cite{Kallosh:1980fi,Howe:1981xy}.}
\be
S_{BR}^{3-loop}\sim \kappa^{4} \int d^4  x   C_{\alpha \beta \gamma \delta} (x)  C_{\dot \alpha \dot \beta \dot \gamma \dot  \delta}  (x) C^{\alpha \beta \gamma \delta}  (x)  C^{\dot \alpha \dot \beta \dot \gamma \dot  \delta}  (x) \ .
\label{BR}\ee
The $SU(8)$ covariant form of the 3-loop counterterm is based on the  linearized superfield of \N=8 supergravity
\be
W_{ijkl} (x, \theta, \bar \theta)= {1\over 4!} \epsilon _{ijklmnpr} \overline W^{mnpr}  (x, \theta, \bar \theta) \ .
\label{self}\ee
The first component is a scalar superfield. The \N=8 supersymmetric, $SU(8)$ invariant generalization of the square of the
Bel-Robinson tensor  is given by a superaction: a superinvariant which is an integral over even-dimensional submanifold of the full superspace. There is a unique  superinvariant :
\be
S^{3}_{UV}=A_3 {\kappa^{4}\over \epsilon} \int d^4 x \, D^{[i_1...i_4][j_1...j_4]} \bar D^{[k_1...k_4][l_1...l_4]} \times L_{ i_1...i_4, j_1...j_4,k_1...k_4, l_1...l_4} \ .
\label{superaction}\ee
Here
\be
D^{[i_1...i_4][j_1...j_4]}\equiv D^{[i_1}_{(\alpha_1 }\dots D^{i_4]}_{\alpha_4) }D^{[j_1}_{(\beta_1 } \dots D^{j_4]}_{\beta_4 )} \epsilon^{\alpha_1 \beta_1} \dots
\epsilon^{\alpha_4 \beta_4} \ .
\ee
The kernel is a $SU(8)$ tensor corresponding to a square Young tableaux
\be
L_{ i_1...i_4, j_1...j_4,k_1...k_4, l_1...l_4}= (W_{ i_1...i_4}  W_{ j_1...j_4}  W_{ k_1...k_4}  W_{ l_1...l_4} )_{\underline{232848}}
\ee
It is in the $\underline{232848}$ representation of the $SU(8)$.
This is a structure of the UV divergence,  with $\epsilon= d-4$,  at the 3-loop order in \N=8 supergravity.

This invariant could have been a reason for the 3-loop logarithmic divergence
if the coefficient $A_3$ in front of it would not vanish. The actual computation in \cite{Bern:2007hh} shows that $A_3=0$.
However,
by inserting  ten or more space-time derivatives, one can construct the  linearised candidate  counterterms starting from the 8-loop order, and we do not know what are the coefficients in front of all of them.

The bosonic part of an infinite number of UV countertems quartic in the Weyl curvature  spinor $ ( C_{\alpha \beta \gamma \delta}, \;  \bar C^{\dot \alpha \dot \beta \dot \gamma \dot  \delta}) $
for arbitrary loop order $L$ is of the general form
\be
S^{L}= A_L {\kappa^{2(L-1)}\over \epsilon}  \int d^4 x\,  C_{\alpha \beta \gamma \delta} (x)\, C^{\alpha \beta \gamma \delta} (x)\, \partial ^{2(L-3) } \bar C_{\dot \alpha \dot \beta \dot \gamma \dot  \delta}\, \bar C^{\dot \alpha \dot \beta \dot \gamma \dot  \delta} \ .
\label{SL}\ee
The derivative term $\partial ^{2(L-3) }$ is symbolic; it means various possible insertions of the covariant derivatives acting on each of the Weyl spinors.  These terms can be presented in a manifestly \N=8 supersymmetric form  with the hidden $E_{(7,7)}$ symmetry, as we have shown in case $L=8$ in eq. (\ref{S8}). At the linearized level it will correspond to $\partial ^{10 }$ insertion into Weyl spinors.
And, of course, many other geometric invariants with higher powers of superfields are possible.

The 4-graviton expression in eq. (\ref{BR}) follows from the superaction (\ref{superaction}) in a simple way: each set of 4 derivatives acts on the superfield with the same set of $SU(8)$ indices, e.g.
\be
(D^{[i_1...i_4] } W_{i_1...i_4})( D^{ [j_1...j_4]} W_{j_1...j_4}) \sim  C_{\alpha \beta \gamma \delta}   C^{\alpha \beta \gamma \delta} \ ,
\ee
and the analogous expression for the conjugate ones.
This gives a product of four $SU(8)$ singlets which is a 4-graviton amplitude. One can see from this procedure that the gravity part of the counterterm is unique, just one particular arrangement of 16 fermionic derivatives.

The situation with scalars is significantly more complicated. We are looking at an $SU(8)$ tensor $M_4(1^{i_1...i_4}, 2^{j_1...j_4}, 3^{k_1...k_4} , 4^{k_1...k_4})$ with $4\times 4$ indices. The scalar terms in the superfield enter as the first component, as well as many other terms with the space-time derivatives. The 16 fermionic derivatives will hit the 4 superfields and the 4-scalar amplitude will have many entries. Independently of the space-time derivatives which do not carry the $SU(8)$ indices, one can imagine two different  $SU(8)$ structures. For example, there are structures where two scalars form an $SU(8)$ tensor and two other scalars form another $SU(8)$ tensor, as only indices of the 1st and the second scalar (and the 3d and the 4th) form the 8-component $SU(8)$  $\epsilon$-tensor,
\be
M_4^I(1^{i_1...i_4}, 2^{j_1...j_4}, 3^{k_1...k_4} , 4^{k_1...k_4}) =A_3\, {\kappa^{4}\over \epsilon}\, \epsilon^{i_1i_2 i_3 i_4 j_1 j_2 j_3 j_4}\,  \epsilon^{k_1k_2 k_3 k_4l_1l_2 l_3l_4}\, f^I(p^i) \ .
\label{I}\ee
A different structure may occur when all 4 particles are interconnected, for example
\be
M_4^{II}(1^{i_1...i_4}, 2^{j_1...j_4}, 3^{k_1...k_4} , 4^{k_1...k_4}) = A_3\, {\kappa^{4}\over \epsilon}\, \epsilon^{i_1i_2 i_3 i_4 j_1j_2 k_3 k_4}\,  \epsilon^{k_1k_2 j_3 j_4 l_1l_2 l_3 l_4}\, f^{II}(p^i) \ .
\label{II} \ee
Here the functions $f^I(p^i)$  and $ f^{II}(p^i)$ depend on particle  momenta.
Each of the  8-component $SU(8)$ $ \epsilon$-tensors has a mix of the 1st, 2d and 3d particles, or a mix of  the 2d, 3d and 4th particles. These two different structures  are definitely present in the 4 scalar amplitude \cite{Tomas}.

Here we will give a simple argument why both structures necessarily should be present. In \N=8 QFT we may compute all tree diagrams which contribute to the 4 scalar amplitude. We should use for this purpose a scalar-scalar-graviton cubic coupling and look for the graviton exchange. The corresponding vertex is given by
\be
\sqrt g \partial_\mu  \phi _{ijkl} \partial_\nu \bar \phi ^{ijkl} h^{\mu\nu} \ .
\ee
 A diagram of this kind will produce the $SU(8)$ structure I given in eq. (\ref{I}) with account of the self-duality of the scalar given in eq. (\ref{self}). In addition, there are  contact terms with 4 scalars, which were constructed from the requirement of unbroken $E_{(7,7)}$, see  \cite{Cremmer:1979up,Kallosh:2008ic}. The scalars are in the ${E_{(7,7)}\over SU(8)}$ coset space. In the unitary  gauge, in which only 70 scalars are present, the kinetic term of scalars produces the  $SU(8)$ covariant 4-scalar coupling. The relevant contact term is \cite{Cremmer:1979up,Kallosh:2008ic}
 \be
 \partial_\mu \phi _{ijkl} \partial^ \mu \bar \phi ^{klmn} \phi_{mnpq} \bar \phi^{pqij} \ .
 \ee
This contact term is the source of the structure II.

Here we have to add that the relation between the tree order \N=8 amplitudes and the 3-loop counterterm is simple for the 4-particle amplitude: they are related by a multiplication of the tree amplitude by the $stu$ factor, where $s, t, u$ are the Mandelstam variables for the 4-particle amplitude.

Using helicity formulation for \N=8 supergravity \cite{Bern:1998ug}, one can describe the 4-graviton tree amplitude as follows,
\be
M^{tree}_4(1^- ,  2^-,  3^+,  4^+)= {i\over  \kappa^2} \, {\langle 12 \rangle^4 [34]^4\over stu}\ ,
\label{M4tree}\ee
in agreement with the string theory computation in \cite{Green:1982sw}.
Here we have adapted the answer in eq. (2.24) of \cite{Bern:1998ug} to the form of  eq. (\ref{M4tree}) above which is more suitable for us, we have used various identities of the helicity formalism. The one-loop finite 4-graviton amplitude is proportional to
\be
M^{1-loop}_4(1^- ,  2^-,  3^+,  4^+)\sim {\langle 12 \rangle^4 [34]^4}\times f_{box} (p_1, p_2, p_3, p_4) \ .
\ee
Here $f_{box} (p_1, p_2, p_3, p_4)$ is a box diagram. The manifestly supersymmetric version of the finite one-loop \N=8 diagrams is given in \cite{Kallosh:2007ym}. Here we are interested in the 3-loop counterterm. To provide manifest supersymmetry, the pure gravitational part of it must be given by
\be
M^{3-loop}_4(1^- ,  2^-,  3^+,  4^+) \sim {\kappa^{4}\over \epsilon} {\langle 12 \rangle^4 [34]^4}\sim  {\kappa^{4}\over \epsilon} stu \, M^{tree}_4(1^- ,  2^-,  3^+,  4^+) \ .
\label{M43loop}
\ee
One can see this also from the Bel-Robinson part of the 3-loop counterterm in eq. (\ref{BR}). One can use  {\it the dictionary between the superfields and  helicity formalism}. It  is based on  `stripping' of Lorentz indices from the curvature tensor\footnote{This is in spirit of  the  recent derivation of the Nair construction for \N=4 Yang Mills theory in \cite{Drummond:2008vq} where the    `stripping' of Lorentz indices was applied to the vector field strength spinor. We have checked that the analogous  `stripping' of Lorentz indices procedure leads to a derivation of the generating functional of the tree amplitudes in \N=8 supergravity proposed in \cite{Bianchi:2008pu}.}
\be
C_{\alpha \beta\gamma\delta}(p) \Rightarrow \lambda_{\alpha}(p) \lambda_{\beta}(p) \lambda_{\gamma}(p) \lambda_{\delta}(p) h(p) \ , \qquad \bar C_{\dot \alpha \dot \beta\dot \gamma\dot \delta}(p) \Rightarrow  \bar \lambda_{\dot \alpha}(p) \bar \lambda_{\dot\beta}(p) \bar \lambda_{\dot \gamma}(p) \bar \lambda_{\dot\delta}(p) \bar h(p)\ .
\label{dictionary}\ee
With this dictionary at hand we see that the superfield version of the 3-loop counterterm means the following UV divergent 4-graviton amplitude
\bea
M^{3-loop}_4(1^- ,  2^-,  3^+,  4^+) h(p_1) h(p_2) \bar h(p_3) \bar h_(p_4) & \sim &{\kappa^{4}\over \epsilon}     C_{\alpha \beta \gamma \delta} (p_1)  C_{\dot \alpha \dot \beta \dot \gamma \dot  \delta}  (p_2) C^{\alpha \beta \gamma \delta}  (p_3)  C^{\dot \alpha \dot \beta \dot \gamma \dot  \delta}  (p_4) \nonumber\\
 &\sim & {\kappa^{4}\over \epsilon}  \langle 12 \rangle^4 [34]^4 h(p_1) h(p_2) \bar h(p_3) \bar h_(p_4) \ .
\label{BRhelicity}\eea
This is a shortcut for the procedure of replacing each  curvature tensor $C_{\mu\nu\lambda\delta} $ by a linear in the graviton expression of the type $\partial_\mu \partial_ \lambda  h_{\nu\delta}$.  To get the two physical states out of the ten-component graviton  one has to replace $h_{\nu\delta}$ by the product of polarization tensors $\epsilon_\nu\epsilon_\delta$. To pick up from this only two physical states one has to impose restrictions on the polarization vectors.  For a  nice pedagogical explanation with examples of the transition from  the well known but not efficient Feynman rules of QFT with polarization tensors to the helicity formalism  see  \cite{Dixon:1996wi}. The results for the 4-graviton amplitude in QFT procedure are given in eq. (\ref{BRhelicity}) but they are obtained in a simple way using the dictionary in eq. (\ref{dictionary}) .

This means that the $SU(8)$ structure of the 3-loop counterterm is the same as the tree amplitudes since the multiplication by the $stu$ factor does not affect the $SU(8)$ structure.

\section{Light-cone counterterms}

 The  chiral superfield $\phi(x, \theta, \bar \theta )$ and the anti-chiral superfield $\phi(x, \theta, \bar \theta )$ both have dimension -2 and satisfy the constraints;
\be
d^m \phi=0\ , \qquad \bar d_m \bar \phi =0 \ .
\ee
Here
\be
d^m \equiv- {\partial \over \partial\bar \theta _m}- {i\over \sqrt 2} \theta^m\partial^+\ , \qquad \bar d_m \equiv  {\partial \over \partial \theta ^m}+ {i\over \sqrt 2} \bar \theta_m\partial^+ \ .
\label{der}\ee
The anti-chiral superfield $\phi(x, \theta, \bar \theta)$  depends only on $ \theta$ in the corresponding basis.
Note that the anti-chiral field is not an independent one as they both describe  a   single CPT invariant multiplet. The relation between them is given by
$
\phi= {1\over (\partial ^{+}) ^4} d^1 ...d^8 \bar \phi
$.
The measure  of integration has dimension $-4+ 8=4$.
Consider, for example,  the  3-loop counterterm:
\be
B_3 {\kappa^4\over \epsilon}  \delta^4 (p_1+p_2+ p_3 +p_4) \int d^8 \theta \, d^8 \bar \theta  \, \phi(p_1, \theta, \bar \theta)\, \phi(p_2, \theta, \bar \theta)\, \bar  \phi(p_3, \theta, \bar \theta)\, \bar \phi(p_4, \theta, \bar \theta)\, f(s, t, u)\ .
\label{covariant}\ee
Here $f(s,t,u)$ is a polynomial in particle momenta of dimension +8, for example, $f= t^2 u^2$.
To increase the loop number we will have to insert various derivatives. This will permit  the more general coupling dependence $\kappa^{2(L-1)}$, and the function $f$ will be of dimension $2L+2$.

To perform the Grassmann integration we note that when only scalars are present, the light-cone superfield simplifies, namely
\be
\phi(p, \theta, \bar \theta) = -e^{  \theta^m \bar \theta_m p^+}\, \theta^{mnpq} \phi_{mnpq}(p) \ ,  \qquad \bar \phi(p, \theta, \bar \theta) = - e^{  \theta^m \bar \theta_m p^+}\bar \theta_{mnpq} \bar \phi^{mnpq}(p) \ .
\label{4powers} \ee
 Only the terms quartic in Grassmann variables in each superfield contribute to (\ref{covariant}), there is no contribution from the exponents. The answer is simple:
\be
{\kappa^4\over \epsilon}  \delta^4 (p_1+p_2+ p_3 +p_4)\, t^2\, u^2\, \bar \phi^{ijlk}(p_1)\, \phi_{ijkl}(p_2)\, \bar \phi^{mnpq}(p_3)\, \phi_{mnpq}(p_4) \ .
\label{4scalars}\ee
This is a Lorentz covariant answer for the 4-scalar amplitude which we find in the light-cone gauge. It is in  agreement with the $SU(8)$ structure which we called $M^I_4$  in the previous section.

We may now also consider some other parts of the integral (\ref{covariant}), for example, the  4-graviton amplitudes. We find
\be
{\kappa^4\over \epsilon}  \delta^4 (p_1+p_2+ p_3 +p_4)\, t^2\, u^2\, h(p_1)\, h(p_2)\, \bar h (p_3)\, \bar h (p_4) f'(p^+_i) \ .
\label{4gr}\ee
where $f'(p^+_i)$ is some non-polynomial function of $p^+_i$. There are two sources of this $p^+$ dependence: first there is a dependence on $p^+$ in front of $h$ and $\bar h$ helicity states of the graviton in the superfield. Another source is coming from rewriting the chiral and anti-chiral superfields in the common basis. This involves $\partial^+$ operator via the spinorial derivatives. Expression in eq. (\ref{4gr}) has to be compared with the covariant expression (\ref{BRhelicity}).

In the light-cone gauges one can look for the helicity brackets $\langle ij \rangle$ and $[ij]$ in the form \cite{Mangano:1990by}, \cite{Dixon:1996wi}.
\be
\langle ij \rangle = {\tilde p_i p_j^+ - \tilde p_j p_i^+ \over \sqrt{p_i^+ p_j^+}} \ , \qquad  [ji]= {\tilde {\bar  p}_i p_j^+ - \tilde{\bar  p}_j p_i^+ \over \sqrt{ p_i^+ p_j^+}}\ ,  \qquad \tilde p_i\equiv p_i^1+ip_i^2 \qquad \tilde {\bar p}_i \equiv p_i^1- ip_i^2
\ee
 However, in eq. (\ref{4gr}) there are no transverse derivatives $\tilde \partial, \tilde {\bar \partial}$ which could supply $\tilde p, \tilde{\bar p}$. Therefore the gravitational part of the integral (\ref{covariant}) is not satisfactory.

To get the second structure in the scalar amplitudes, $M^{II}
_4$, we have to replace some of the space-time covariant derivatives in the function $f$ by the bilinear  combination of the fermionic derivatives, which carry an $SU(8)$ index. For example, any $p_{ i \mu}\cdot p^{\mu}_j$ can be replaced by some set of covariant derivatives defined in eq. (\ref{der}). One possibility would be to use
\be
C_3 {\kappa^4\over \epsilon}  \delta^4 (p_1+p_2+ p_3 +p_4) \int d^8 \theta \, d^8 \bar \theta  \, \phi(p_1, \theta, \bar \theta) \bar d_{mnpq} \phi(p_2, \theta, \bar \theta) d^{mnpq}\bar  \phi(p_3, \theta, \bar \theta) \bar \phi(p_4, \theta, \bar \theta) g(s, t, u, p^+_i)
\ee
The local counterterms in the light-cone gauges may have some inverse, non-polynomial dependence on $p^+_i$ but are expected to have only polynomial dependence on $s,t,u$. After Grassmann integration we may find the 4-scalar amplitude required by the covariant counterterm. However, when we look at the 4-graviton part of the integral, we cannot reproduce the required helicity structure due to absence of $p_i, \bar p_i$-dependence. We find therefore that the covariant and the light-cone counterterms do  not match and therefore
\be
A_3=B_3=C_3=0
\ee
We conclude that the divergences in the 3d loop  in different gauges are incompatible and therefore should not appear in actual computations. This may be one of possible explanations of the  results of   \cite{Bern:2007hh}.

\section{Counterterms for the $n$-particle S-matrix for $n>4$}
The $n< 4$ on-shell amplitudes vanish in \N=8 supergravity, the case $n=4$ we studied above. For the n-particle amplitudes  with $n>4$,  we may again compare the Lorentz covariant linearized counterterms with the light-cone ones.
In the covariant approach we would expect for the all-scalar amplitudes
\be
A_L(n)  {\kappa^{2L-2}\over \epsilon} \delta^4 (p^1+p^2+...  p^n)\, f(p^i)\, \phi(p^1) \, \phi(p^2) \dots \bar \phi(p^n)
\ee
with some contraction of the $SU(8)$ indices on the scalars $\phi_{ijkl}$ and some covariant function $f(p^i)$ which has a particular weight.

To construct the $n>4$  scalar amplitude in the light-cone superspace  one would start with the integral over $n$ superfields and only covariant space-time derivatives as well as spinorial derivatives. One should not have $\partial, \, \bar \partial $ derivatives since the all-scalar amplitude should be covariant. The same issue as before arises, namely, how the helicity amplitudes for vectors and gravitons can come out as a result of the integral over the light-cone superspace?

Thus  we do  not see how to build  the light-cone counterterms in \N=8 supergravity for $n$-particle amplitudes which would survive the comparison between the covariant and the light-cone gauges.

\section{All-loop argument for the $n$-particle UV finiteness}
The most striking feature of the argument about incompatibility of the UV divergences in different gauges is that it refers to any loop order  for the  S-matrix. The reason is the following. The argument is based on the use of the linearized superfields only (equivalent to the use of global \N=8 supersymmetric Ward identities). In such case, a simple insertion of derivatives will increase the dimension of the counterterms and require higher powers of $\kappa$. This will be the only difference between higher loop counterterms and the 3-loop counterterm for the 4-particle S-matrix.  Analogously,  the difference between the  lowest loop order and  the  higher loop orders for any  $n>4$ matrix elements will be just  the insertion of derivatives. In the 4-particle case, the simple space-time derivatives are  $SU(8)$ neutral and Lorentz covariant and will not change the discrepancy between the UV divergences in different gauges, which we described here in all details for the case of the 3 loop UV divergence in the 4-scalar amplitude. For higher $n$, the insertion of additional covariant derivatives will not remove the discrepancy with Lorentz non-covariant terms which we discussed above.

An important
 condition, which must be satisfied to make our argument about the all-loop finiteness of \N=8 supergravity valid, is the absence of anomalies. An equivalence theorem of the type proved in \cite{Kallosh:1972ap} for non-Abelian gauge theories is based on a possibility to change the variables in the  path integral so that one can reach different  gauges via the change of variables.\footnote{ It is more difficult to generalize to supergravity the well known combinatorial proof of gauge independence in non-Abelian gauge theories \cite{'tHooft:1972ue}.}  The equivalence theorem \cite{Kallosh:1972ap} takes place  for the S-matrix. The possibility to perform the relevant change of variables in the path integral  is associated with the generalized BRST  symmetry of the gauge-fixed action. In supergravity, at least in \N=1 case, the corresponding Feynman rules have been derived in \cite{Kallosh:1977ik}.  They agree with the supergravity Feynman rules derived by canonical quantization in \cite{Fradkin:1977wv}. These rules are rather non-trivial due to the open algebra of the local supersymmetry generators, which closes only on shell. The features of these Feynman rules include 4-ghost coupling and the third ghost in addition to the Faddeev-Popov ghosts required in non-Abelian gauge theories.  It is plausible that  an analogous generalization of the BRST symmetry can be provided for \N=8 supergravity. On the other hand, it appears that the unitarity method of \cite{Bern:1994zx} leads directly to gauge-independent answers for the S-matrix.

 Having established the BRST symmetry of the path integral, one can hope that the formal BRST symmetry remains valid in perturbation theory, which requires  the absence of anomalies. In such case the S-matrix should be gauge independent. Our observation that the  UV divergent terms in the S-matrix in different  gauges are incompatible  with each other  leads to a {\it prediction that the perturbative \N=8 supergravity  amplitudes are all-loop finite,  if there are no anomalies}.\footnote{The recent proof of the no-triangle conjecture in \cite{BjerrumBohr:2008ji}, \cite{ArkaniHamed:2008gz}
 may be an indication of the absence of anomalies as suggested in \cite{Kallosh:2007ym}. However, we will need to know more to establish  the absence of anomalies to all loop order.}

\section{$SU(8)$ and \E , fermions, vectors and one-loop triangle anomalies}
The local $SU(8)$ symmetry is chiral, therefore the standard triangle one-loop anomalies are possible,  a priori. These anomalies have been in fact computed in \cite{Marcus:1985yy} using the methods developed in \cite{AlvarezGaume:1983ig}, \cite{Alvarez:1984yi}. The $SU(8)$ anomalies in the first loop get a contribution from spin 1/2, spin 3/2 as well as from the chiral vector  of spin 1. The chiral fermion contribution  does not cancel by itself,  the cancellation is due to the account of the chiral nature of the vectors of \N=8 supergravity.

The action is given in terms of the 28 real Abelian vector fields  ${\cal A}_\mu^{IJ}$. They  enter only via the $U(1)$ gauge invariant Maxwell field strength
$
F_{\mu\nu}^{IJ}= \partial_\mu  {\cal A}_\nu^{IJ}- \partial_\nu  {\cal A}_\mu^{IJ}
$.
However, these vector fields in the lagrangian do not describe helicity $\pm1$ states since there is no real 28-dimensional representation in $SU(8)$ and moreover, the fields ${\cal A}_\mu^{IJ}$ do not transform under the local $SU(8)$ transformations, they transform under local $U(1)$ and global \E. However, there exist a  the combination of these vectors  with the scalars which correspond to physical helicity $\pm 1$ states.

First we define the self-dual and anti-self dual field strengths
\be
F^{+}_{\mu\nu IJ}= {1\over 2} (F_{\mu\nu}^{IJ}+  i \tilde  F_{\mu\nu}^{IJ})\qquad,\qquad
F^{-IJ}_{\mu\nu}= {1\over 2} (F_{\mu\nu}^{IJ}-  i \tilde  F_{\mu\nu}^{IJ})\ ,
\ee
in terms of the original field strength $F_{\mu\nu}^{IJ}$.
They are related by complex conjugation $(F^{+}_{\mu\nu IJ})^*= F_{\mu\nu}^{-IJ}$, the dual field strength $\tilde{F}_{\mu\nu}^{IJ}$ being  defined  as
$
 i \tilde F_{\mu\nu} \equiv  {1\over 2} \epsilon_{\mu\nu\rho\sigma} F^{\rho\sigma  IJ}\ .
$
The next step involves to use a doublets in \E \,
\begin{eqnarray}\label{F1F2}
F^{+}_{1IJ}\equiv\frac{1}{2}(G^{+}_{IJ}+F^{+}_{IJ})\ \ \ ,\ \ \
F^{+IJ}_{2}\equiv\frac{1}{2}(G^{+}_{IJ}-F^{+}_{IJ})\ ,
\end{eqnarray}
which require to combine the field strength above with the derivatives of the Lagrangian over these fields
\begin{eqnarray}\label{G+}
G^{+}_{IJ}\equiv-\frac{4}{e}\frac{\delta {\cal{L}}}{\delta F^{+}_{IJ}}\ .
\end{eqnarray}
Now one can get the $SU(8)$ transforming vector fields by contracting the \E \, doublet with the scalar-dependent 56-bein
\begin{eqnarray}\label{gauge}
\cV=\left(
                                        \begin{array}{cc}
                                          u_{ij}{}^{IJ} & v_{ijKL} \\
                                          v^{klIJ} & u^{kl}{}_{KL} \\
                                        \end{array}
                                      \right)\ .
\end{eqnarray}
The rigid  $E_{7(7)}$ symmetry is realized on scalars as a multiplication from the right, whereas the  local $SU(8)$ transformation acts on the 56-bein from the left and is completely independent of the $E_{7(7)}$ transformation.
The $SU(8)$ covariant vector field strength $ \bar F^{+}_{\mu\nu ij} $ is defined via the multiplication of the 56-bein on the \E \, doublet
\begin{eqnarray}\label{barF}
\cV \, \left(
  \begin{array}{c}
    F^{+}_{1\mu\nu} \\
    F^{+}_{2\mu\nu} \\
  \end{array}
\right)= \left(
  \begin{array}{c}
    \bar F^{+}_{\mu\nu ij} \\
   \O^{+}_{\mu\nu}{}^{ij} \\
  \end{array}
\right)\ .
\end{eqnarray}
Here $ \O^{+}_{\mu\nu}{}^{ij}$ depends on fermions. Thus the  $\mathbf{28}$ $SU(8)$ self-dual vector is
\be
 \bar F^{+}_{\mu\nu ij}= u_{ij}{}^{IJ} F^+_{1\mu\nu IJ} + v_{ijIJ} F_{2\mu\nu}{}^{+IJ}
\ee
The anti-self-dual $\overline {\mathbf{28}}$ in $SU(8)$ is a conjugate one. It is this vector field strength which appears in the local supersymmetry of fermions, for example
\be
\delta_{susy}\chi ^{ijk} = 3\sigma^{\mu\nu} \bar F^{-[ij} _{\mu\nu} \epsilon^{k]} +...
\ee
It is a complicated function of the original vector field in the Lagrangian and scalar fields and it transforms under local $SU(8)$ before the gauge-fixing $SU(8)$ and does not transform under the global  \E \,. After gauge-fixing it transforms under \E \, {\it non-linearly},   same as as the fermions, via the compensating $SU(8)$ transformation \cite{Kallosh:2008ic} depending on the parameters of the global  \E \,.

Thus we have provided here the justification of the non-trivial contribution of the chiral vectors to the BRST symmetry part associated with the local $SU(8)$ symmetry despite the fact that the vector fields in the Lagrangian do not transform under local $SU(8)$.

\subsection{Anomaly counting }

In four dimensions helicity $h$ state contribution to anomaly is proportional to
\be
  (-)^{2h} \, 3h\,  \rm tr \, F\wedge F\wedge F
\label{hel}\ee
Here $\rm tr \, F\wedge F\wedge F$ is a pure ``gluon'' six form associated with the $SU(8)$ anomaly of \N=8 supergravity.
This means that the contribution to anomaly from $\mathbf{8}$, $\mathbf{28}$ and $\mathbf{56}$ (gravitini, vectors and graviphotini, respectively) is defined via
\be
\rm str \, \lambda^a \lambda ^b \lambda^c= d^{abc} C_3
\ee
Here the Casimir $C_3$ has the ratios $1:4:5$ for $\mathbf{8}$, $\mathbf{28}$ and $\mathbf{56}$.  With account of eq. (\ref{hel}) the total contribution to $SU(8)$ anomaly vanishes \cite{Marcus:1985yy}:
\be
(-3)\times 1+ (2) \times 4 +(-1) \times 5=0
\ee
 \subsection{Implications from the vanishing $SU(8)$ anomaly for anomalies in  \E \, symmetry and supersymmetry}

 When the local $SU(8)$ symmetry is gauge-fixed one has a global \E \, symmetry acting on various fields. One of the important properties of the classical \N=8 supergravity is that the \E \, symmetry is continuous and therefore it is associated with the conserved Noether current, $\partial_\mu j^\mu=0$, \cite{Kallosh:2008ic}. The non-perturbative effects break this symmetry to a discrete one which means that there is no conserved current.

 The fact that the $SU(8)$ one-loop triangle anomaly vanishes may be an indication that there is no anomaly in \E \, symmetry. The 70 generators of ${G\over H}$ symmetry, $X$ form an algebra where the 63 $T$-generators form the maximal  $SU(8)$ subalgebra, $[T, T]\sim T$ and  $[X, T]\sim X$ and $[X, X]\sim T$. This means that the Wess-Zumino  consistency condition for anomalies  \cite{Wess:1971yu} suggests that the total $G=$\E \, may not be anomalous, at least at the one-loop level. The consistency condition requires a constraint between anomalies in case the corresponding symmetries form an algebra:
 \be
 \delta_{\Lambda_1} G(\Lambda_2) -  \delta_{\Lambda_2} G(\Lambda_1) = G(\Lambda),  \qquad  \Lambda = [\Lambda_1, \Lambda_2]
 \ee
 Here the anomaly is $G(\Lambda)= {\delta \over \delta \Lambda} \Gamma$ and $\Gamma$ is an effective action. Since the $SU(8)$ is a subalgebra, if it would be anomalous it would mean that also the coset part of \E \, symmetry would have to be anomalous. But since the $SU(8)$ is not anomalous, it is consistent with the non-anomalous total  \E \, symmetry
 \footnote{For the 4-point 1-loop amplitudes the absence of \E \, anomaly is shown in \cite{Tomas}.}. It will be interesting to find out the general status of these anomalies in \N=8 SG.

 A strategy of understanding anomalies may include the studies of the 70 shift symmetry generators of \E \,. This non-anomalous symmetry means, in particular, suggests that there is a  moduli space for scalars in perturbative theory. This is in a sharp contrast with the non-perturbative effects in \N=8 supergravity. For example, in  the extremal black hole solutions scalars near the horizon  are  attracted to a particular value defined by the black hole charges, the moduli space is lifted. Hopefully it will be possible to understand all of this better and find out if the unbroken \E \, symmetry is valid in perturbative QFT and if it helps with the studies of the UV properties of \N=8 SG.

\vskip 2 cm

We should emphasize that any statement about the all-loop finiteness of \N=8 supergravity should not be taken lightly. The theory is extremely complicated when the starting point is the Lagrangian and the Feynman rules\footnote{It is also possible that from the new perspective it is the simplest QFT ever  \cite{ArkaniHamed:2008gz}.}. It might happen, for example,  that a more detailed investigation will reveal some subtleties of the UV divergences in the light-cone gauges which  remove the discrepancy with the covariant  candidate divergences or violate the equivalence theorem.  One can consider this article as an invitation to construct the possible  light-cone superinvariants and compare them with the covariant ones,  to study the presence/absence of various anomalies,  and either confirm or disprove the conclusions of the present paper. It has been suggested recently in \cite{Stelle} that is should be possible to construct the light-cone counterterms by converting the covariant ones into the light-cone candidates of the UV divergences. It would be interesting to see the realization of this program.

\section*{Acknowledgments}

I thank   L.~Alvarez-Gaume, L. Brink, M. Bianchi, S. Deser,  H. Elvang, S. Ferrara, D. Freedman, P. Howe, T. Kugo,   Ching Hua Lee, A. Linde, U. Lindtsrom, M. Noorbala, Sung-Soo Kim,  M. Soroush, K. Stelle,    T. Rube and P. Vanhove for the most useful discussions of \N=8 SG.  I am particularly grateful to Z. Bern and   L. Dixon for explaining to me their work.
This work is
supported by the NSF grant 0756174.



\begin{thebibliography}{10}


\bibitem{Cremmer:1979up}
  E.~Cremmer and B.~Julia,
``The SO(8) Supergravity,''
  Nucl.\ Phys.\  B {\bf 159}, 141 (1979);
  B.~de Wit and H.~Nicolai,
 ``N=8 Supergravity,''
  Nucl.\ Phys.\  B {\bf 208}, 323 (1982).



\bibitem{Bern:2006kd}
  Z.~Bern, L.~J.~Dixon and R.~Roiban,
  ``Is N = 8 supergravity ultraviolet finite?,''
  Phys.\ Lett.\  B {\bf 644}, 265 (2007)
  [arXiv:hep-th/0611086].
  M.~B.~Green, J.~G.~Russo and P.~Vanhove,
  ``Ultraviolet properties of maximal supergravity,''
  Phys.\ Rev.\ Lett.\  {\bf 98}, 131602 (2007)
  [arXiv:hep-th/0611273].

\bibitem{Bern:2007hh}
  Z.~Bern, J.~J.~Carrasco, L.~J.~Dixon, H.~Johansson, D.~A.~Kosower and R.~Roiban,
  ``Three-Loop Superfiniteness of \N=8 Supergravity,''
  Phys.\ Rev.\ Lett.\  {\bf 98}, 161303 (2007)
  [arXiv:hep-th/0702112].



\bibitem{Kallosh:1980fi}
  R.~E.~Kallosh,
``Counterterms in extended supergravities,''
  Phys.\ Lett.\  B {\bf 99} (1981) 122;
  R.~Kallosh,
``Counterterms In Extended Supergravities ,''
 Lectures given at Spring School on Supergravity,
Published in Trieste Supergravity School 1981.

\bibitem{Howe:1981xy}
  P.~S.~Howe, K.~S.~Stelle and P.~K.~Townsend,
 ``Superactions,''
  Nucl.\ Phys.\  B {\bf 191}, 445 (1981).

\bibitem{Howe:2002ui}
  P.~S.~Howe and K.~S.~Stelle,
 ``Supersymmetry counterterms revisited,''
  Phys.\ Lett.\  B {\bf 554}, 190 (2003)
  [arXiv:hep-th/0211279.


\bibitem{Howe:2004pn}
  P.~S.~Howe,
 ``R**4 terms in supergravity and M-theory,''
  arXiv:hep-th/0408177.

  \bibitem{Stelle} K.~S.~Stelle, talk at the European network wokshop, Varna 2008.

\bibitem{Galperin:1984av}
  A.~Galperin, E.~Ivanov, S.~Kalitsyn, V.~Ogievetsky and E.~Sokatchev,
 ``Unconstrained N=2 Matter, Yang-Mills And Supergravity Theories In Harmonic
 Superspace,''
  Class.\ Quant.\ Grav.\  {\bf 1}, 469 (1984);
  A.~S.~Galperin, E.~A.~Ivanov, V.~I.~Ogievetsky and E.~S.~Sokatchev,
  ``Harmonic Superspace,''
{\it  Cambridge, UK: Univ. Pr. (2001)}




\bibitem{Kallosh:2007ym}
  R.~Kallosh,
  ``The Effective Action of N=8 Supergravity,''
  arXiv:0711.2108 [hep-th].


\bibitem{Bern:1998ug}
  Z.~Bern, L.~J.~Dixon, D.~C.~Dunbar, M.~Perelstein and J.~S.~Rozowsky,
  ``On the relationship between Yang-Mills theory and gravity and its
 implication for ultraviolet divergences,''
  Nucl.\ Phys.\  B {\bf 530}, 401 (1998)
  [arXiv:hep-th/9802162].

\bibitem{Howe:1980th}
  P.~S.~Howe and U.~Lindstrom,
  ``Higher Order Invariants In Extended Supergravity,''
  Nucl.\ Phys.\  B {\bf 181}, 487 (1981).


\bibitem{Brink:1979nt}
  L.~Brink and P.~S.~Howe,
 ``The \N=8 Supergravity In Superspace,''
  Phys.\ Lett.\  B {\bf 88}, 268 (1979).

\bibitem{Brink:1982pd}
  L.~Brink, O.~Lindgren and B.~E.~W.~Nilsson,
  ``N=4 Yang-Mills Theory On The Light Cone,''
  Nucl.\ Phys.\  B {\bf 212}, 401 (1983).


\bibitem{Brink:2008qc}
  L.~Brink, S.~S.~Kim and P.~Ramond,
``$E_7(7)$ on the Light Cone,''
  JHEP {\bf 0806}, 034 (2008)
  [arXiv:0801.2993 [hep-th]].


\bibitem{Mandelstam:1982cb}
  S.~Mandelstam,
 ``Light Cone Superspace And The Ultraviolet Finiteness Of The N=4 Model,''
  Nucl.\ Phys.\  B {\bf 213}, 149 (1983);
  L.~Brink, O.~Lindgren and B.~E.~W.~Nilsson,
  ``The Ultraviolet Finiteness Of The N=4 Yang-Mills Theory,''
  Phys.\ Lett.\  B {\bf 123}, 323 (1983).
  M.~A.~Namazie, A.~Salam and J.~A.~Strathdee,
 ``Finiteness Of Broken N=4 Superyang-Mills Theory,''
  Phys.\ Rev.\  D {\bf 28}, 1481 (1983).


\bibitem{Howe:1982tm}
  P.~S.~Howe, K.~S.~Stelle and P.~K.~Townsend,
  ``The Relaxed Hypermultiplet: An Unconstrained N=2 Superfield Theory,''
  Nucl.\ Phys.\  B {\bf 214}, 519 (1983).



\bibitem{Kallosh:2008ic}
  R.~Kallosh and M.~Soroush,
 ``Explicit Action of E7(7) on N=8 Supergravity Fields,''
  Nucl.\ Phys.\  B {\bf 801}, 25 (2008)
  [arXiv:0802.4106 [hep-th]].



\bibitem{Deser:1977nt}
  S.~Deser, J.~H.~Kay and K.~S.~Stelle,
``Renormalizability Properties Of Supergravity,''
  Phys.\ Rev.\ Lett.\  {\bf 38}, 527 (1977).

   \bibitem{Tomas} R. Kallosh, C. H. Lee and T. Rube, Work in progress.

\bibitem{Green:1982sw}
  M.~B.~Green, J.~H.~Schwarz and L.~Brink,
``N=4 Yang-Mills And \N=8 Supergravity As Limits Of String Theories,''
  Nucl.\ Phys.\  B {\bf 198}, 474 (1982).




\bibitem{Drummond:2008vq}
  J.~M.~Drummond, J.~Henn, G.~P.~Korchemsky and E.~Sokatchev,
 ``Dual superconformal symmetry of scattering amplitudes in N=4
  super-Yang-Mills theory,''
  arXiv:0807.1095 [hep-th].

\bibitem{Bianchi:2008pu}
  M.~Bianchi, H.~Elvang and D.~Z.~Freedman,
  ``Generating Tree Amplitudes in N=4 SYM and N = 8 SG,''
  arXiv:0805.0757 [hep-th].


\bibitem{Dixon:1996wi}
  L.~J.~Dixon, TASI '95
  ``Calculating scattering amplitudes efficiently,''
  arXiv:hep-ph/9601359.

\bibitem{Mangano:1990by}
  M.~L.~Mangano and S.~J.~Parke,
 ``Multiparton amplitudes in gauge theories,''
  Phys.\ Rept.\  {\bf 200}, 301 (1991)
  [arXiv:hep-th/0509223].




\bibitem{Kallosh:1972ap}
  R.~E.~Kallosh and I.~V.~Tyutin,
  ``The Equivalence theorem and gauge invariance in renormalizable theories,''
  Yad.\ Fiz.\  {\bf 17}, 190 (1973)
  [Sov.\ J.\ Nucl.\ Phys.\  {\bf 17}, 98 (1973)];
  B.~W.~Lee and J.~Zinn-Justin,
``Spontaneously Broken Gauge Symmetries. 4. General Gauge Formulation,''
  Phys.\ Rev.\  D {\bf 7} (1973) 1049.


\bibitem{'tHooft:1972ue}
  G.~'t Hooft and M.~J.~G.~Veltman,
  ``Combinatorics of gauge fields,''
  Nucl.\ Phys.\  B {\bf 50}, 318 (1972).

\bibitem{Kallosh:1977ik}
  R.~E.~Kallosh,
 ``Gauge Invariance In Supergravitation,''
  Pisma Zh.\ Eksp.\ Teor.\ Fiz.\  {\bf 26}, 575 (1977)
  [JETP Lett.\  {\bf 26}, 427 (1977)];
  G.~Sterman, P.~K.~Townsend and P.~van Nieuwenhuizen,
  ``Unitarity, Ward Identities, And New Quantization Rules Of Supergravity,''
  Phys.\ Rev.\  D {\bf 17}, 1501 (1978).
  N.~K.~Nielsen,
 ``Ghost Counting In Supergravity,''
  Nucl.\ Phys.\  B {\bf 140}, 499 (1978).
  R.~E.~Kallosh,
  ``Modified Feynman Rules In Supergravity,''
  Nucl.\ Phys.\  B {\bf 141}, 141 (1978).



\bibitem{Fradkin:1977wv}
  E.~S.~Fradkin and M.~A.~Vasiliev,
  ``Hamiltonian Formalism, Quantization And S Matrix For Supergravity,''
  Phys.\ Lett.\  B {\bf 72}, 70 (1977).





\bibitem{Bern:1994zx}
  Z.~Bern, L.~J.~Dixon, D.~C.~Dunbar and D.~A.~Kosower,
 ``One loop n point gauge theory amplitudes, unitarity and collinear limits,''
  Nucl.\ Phys.\  B {\bf 425}, 217 (1994)
  [arXiv:hep-ph/9403226];
  Z.~Bern, L.~J.~Dixon, D.~C.~Dunbar and D.~A.~Kosower,
  ``Fusing gauge theory tree amplitudes into loop amplitudes,''
  Nucl.\ Phys.\  B {\bf 435}, 59 (1995)
  [arXiv:hep-ph/9409265].


\bibitem{BjerrumBohr:2008ji}
  N.~E.~J.~Bjerrum-Bohr and P.~Vanhove,
  ``Absence of Triangles in Maximal Supergravity Amplitudes,''
  arXiv:0805.3682 [hep-th].

\bibitem{ArkaniHamed:2008gz}
  N.~Arkani-Hamed, F.~Cachazo and J.~Kaplan,
  ``What is the Simplest Quantum Field Theory?,''
  arXiv:0808.1446 [hep-th].

\bibitem{Marcus:1985yy}
  N.~Marcus,
``Composite Anomalies In Supergravity,''
  Phys.\ Lett.\  B {\bf 157}, 383 (1985);
  P.~di Vecchia, S.~Ferrara and L.~Girardello,
``Anomalies Of Hidden Local Chiral Symmetries In Sigma Models And Extended
 Supergravities,''
  Phys.\ Lett.\  B {\bf 151}, 199 (1985).

\bibitem{AlvarezGaume:1983ig}
  L.~Alvarez-Gaume and E.~Witten,
  ``Gravitational Anomalies,''
  Nucl.\ Phys.\  B {\bf 234}, 269 (1984).

\bibitem{Alvarez:1984yi}
  O.~Alvarez, I.~M.~Singer and B.~Zumino,
``Gravitational Anomalies And The Family's Index Theorem,''
  Commun.\ Math.\ Phys.\  {\bf 96}, 409 (1984).

\bibitem{Wess:1971yu}
 J.~Wess and B.~Zumino,
``Consequences of anomalous Ward identities,''
 Phys.\ Lett.\  B {\bf 37} (1971) 95;
  M.~B.~Green, J.~H.~Schwarz and E.~Witten,
``Superstring Theory. Vol. 2: Loop Amplitudes, Anomalies And Phenomenology,''
{\it  Cambridge, Uk: Univ. Pr. ( 1987) 596 P. ( Cambridge Monographs On Mathematical Physics)}


\end{thebibliography}
\end{document}